\documentclass[prb,twocolumn,showpacs]{revtex4}     
\usepackage{epsfig,amsmath} 

\begin{document}

\title{Ordered-Current State of Electrons in Bilayer Graphene} 

\author{Xin-Zhong Yan$^{1}$ and C. S. Ting$^2$}
\affiliation{$^{1}$Institute of Physics, Chinese Academy of Sciences, P.O. Box 603, 
Beijing 100190, China\\
$^{2}$Texas Center for Superconductivity, University of Houston, Houston, Texas 77204, USA}
 
\date{\today}

\begin{abstract}
Based on the four-band continuum model, we study the ordered-current state (OCS) for electrons in bilayer graphene at the charge neutrality point. The present work resolves the puzzles that (a) the energy gap increases significantly with increasing the magnetic field $B$, (b) the energy gap can be closed by the external electric field of either polarization, and (c) the particle-hole spectrum is asymmetric in the presence of $B$, all these as observed by the experiment. We also present the prediction of the hysteresis energy gap behavior with varying $B$, which explains the existing experimental observation on the electric conductance at weak $B$. The large energy gap of the OCS is shown to originate from the disappearance of Landau levels of $n$ = 0 and 1 states in conduction/valence band. By comparing with the existing models and the experiments, we conclude that the OCS is a possible ground state of electrons in bilayer graphene.   
\end{abstract}


\maketitle

\section {Introduction}

The study of bilayer graphene (BLG) is a focused area in the condensed-matter physics because of the potential application of BLG to new electronic devices.\cite{Ohta,Oostinga,McCann,Castro} One of the fundamental subjects is to explore the physics of the ground state of electrons in BLG. A number of experiments \cite{Weitz,Freitag,Velasco,Bao} performed on high quality suspended BLG samples have provided the evidence that the ground state is gapped at the charge neutrality point (CNP). In particular, a recent experiment by Velasco {\it et al.} \cite{Velasco} has observed that (i) the ground state is insulating in the absence of external electric and magnetic fields, with a gap $E_{\rm gap} \approx $ 2 meV that can be closed by a perpendicular electric field of either polarization, (ii) the gap grows with increasing magnetic field $B$ as 
$E_{\rm gap} = \Delta_0+\sqrt{a^2B^2 + \Delta_0^2}$
with $\Delta_0 \approx$ 1 meV and $a \approx$ 5.5 meVT$^{-1}$, and (iii) the state is particle-hole asymmetric. On the other hand, theories have predicted various gapped states, such as a ferroelectric-layer asymmetric state \cite{Min,Nandkishore,Zhang,Jung,MacDonald} or quantum valley Hall state (QVH),\cite{Zhang2} a layer-polarized antiferromagnetic state (AF),\cite{Gorbar} a quantum anomalous Hall state (QAH),\cite{Jung,Nandkishore1,Zhang1} a quantum spin Hall state (QSH), \cite{Jung,Zhang1} and a superconducting state in coexistence with antiferromagnetism (SAF). \cite{Milovanovic} The ferroelectric-layer asymmetric and QAH and QSH states all have been ruled out by the experiment.\cite{Velasco} The SAF state is excluded because the real system is an insulator. The AF state cannot reproduce the gap behavior with varying the magnetic field. Recently, the loop-current state has been studied by numerical diagonalization of an effective mean-field Hamiltonian for a finite size lattice \cite{Zhu} and by analytically solving a two-band continuum model (2BCM). \cite{Yan1} Whether the model of this state agrees with the experimental observations on the electronic properties of BLG remains a question.

In this work, using the four-band continuum model (4BCM) for electrons with finite-range repulsive interactions in BLG, we study the ordered-current state (OCS) at the CNP with a rigorous formalism and compare the results with the experimental observations. The importance of using the 4BCM to describe quantitatively the many-body properties of the electron liquid in the BLG has been stressed by the existing works.\cite{Borghi} We here investigate the gap behavior of the OCS with varying the magnetic field $B$, and the particle-hole asymmetry spectra at finite $B$, and the phase transitions in the electron system in the presence of the electric and magnetic fields. We will show that the puzzles (i)-(iii) of the experimental observations can be resolved by the present model of the OCS. With the same 4BCM, we also study the AF state and show that the AF state is not able to reproduce the experimental result for the gap as a function of the magnetic field. 

\section {four-band continuum model}

The lattice structure of a BLG is shown in Fig. 1. The unit cell of BLG contains four atoms denoted as a$_1$ and b$_1$ on top layer, and a$_2$ and b$_2$ on bottom layer with interlayer distance $d \approx 3.34$\AA. The lattice constant defined as the distance between the nearest-neighbor (NN) atoms of a sublattice is $a \approx 2.4$ \AA~. The energies of intralayer NN [between a$_1$ (a$_2$) and b$_1$ (b$_2$)] and interlayer NN (between b$_1$ and a$_2$) electron hopping are $t \approx$ 2.8 eV and $t_1 \approx$ 0.39 eV, respectively. 

\begin{figure}
\vskip 8mm 
\centerline{\epsfig{file=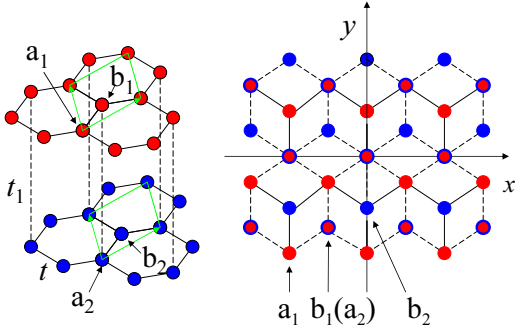,width=7.5 cm}}
\caption{(Color online) Left: Lattice structure of the bilayer graphene. Right: Top view of the bilayer graphene. Atoms a$_1$ (a$_2$) and b$_1$ (b$_2$) are on the top (bottom) layer.} 
\end{figure} 

The first Brillouin zone and the two valleys $K$ and $K'$ in the momentum space are depicted in Fig. 2. For the carrier concentration close to the CNP, we need to consider only the states with momenta close to the Dirac points $K = (4\pi/3,0)$ and $K' = -K$. We here define the operator
$C^{\dagger}_{vk\sigma}=(c^{\dagger}_{a_1,v+k,\sigma},c^{\dagger}_{b_1,v+k,\sigma}
,c^{\dagger}_{a_2,v+k,\sigma},c^{\dagger}_{b_2,v+k,\sigma})$, where $v = K$ or $K'$, $c^{\dagger}_{l,v+k,\sigma}$ creates a spin-$\sigma$ electron of momentum $k$ in valley $v$ of $l$ sublattice, and $k$ is measured from the Dirac point $K$ ($K'$) and confined to a circle $k \leq 1/a$ in $K$ ($K'$) valley. With the operator $C^{\dagger}_{vk\sigma}$, the Hamiltonian describing the noninteracting electrons is given by
\begin{eqnarray}
H_0&=&\sum_{vk\sigma}C^{\dagger}_{vk\sigma}H^0_{vk}C_{vk\sigma} \label {hm}
\end{eqnarray}
with
\begin{eqnarray}
H^0_{vk} &=& \begin{pmatrix}
0& e_{vk}&0&0\\
e^{\ast}_{vk}&0&-t_1&0\\
0&-t_1&0&e_{vk}\\
0&0&e^{\ast}_{vk}&0\\
\end{pmatrix} \label{h0}
\end{eqnarray}
where $e_{vk} = \epsilon_0(s_v k_x+ik_y)$, $s_v = 1$ (-1) for $k$ in the valley $K$ ($K'$), and $\epsilon_0 = \sqrt{3}t/2$. We hereafter use the units of $\epsilon_0$ = 1 and $a$ = 1. 

The interaction Hamiltonian is
\begin{eqnarray}
H' = U\sum_{lj}\delta n_{lj\uparrow}\delta n_{lj\downarrow}+\frac{1}{2}\sum_{li\ne l'j}v_{li,l'j}\delta n_{li}\delta n_{l'j} \label {int}
\end{eqnarray}  
where $\delta n_{li\sigma}$ is the number deviation of electrons with spin $\sigma$ from its average occupation at site $i$ of sublattice $l$ (hereafter denoted as $li$ for short), $\delta n_{li}=\delta n_{li\uparrow}+\delta n_{li\downarrow}$, $U$ is the on-site interaction, and $v_{li,l'j}$ is the interaction between electrons at sites $li$ and $l'j$. Within the mean-field approximation (MFA), since the interaction $v_{li,l'j}$ appears in the exchange self-energy, it can be considered as a finite-range interaction by taking into account the screening effect due to the electronic charge fluctuations.\cite{Yan} The total Hamiltonian $H_0 + H'$ satisfies the particle-hole symmetry. \cite{Yan}

\begin{figure}
\vskip 8mm 
\centerline{\epsfig{file=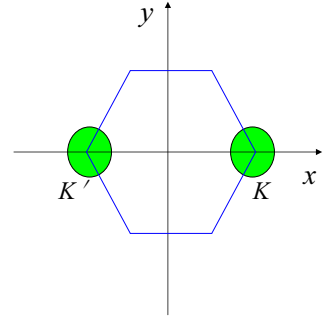,width=5. cm}}
\caption{(Color online) The first Brillouin zone and the two valleys $K$ and $K'$.} 
\end{figure} 

\section {ordered-current state}

In the ordered-current state for which there is no antiferromagnetism, the effective interaction under the MFA is given by
\begin{eqnarray}
H' &\approx& \sum_{li\ne l'j\sigma}v_{li,l'j}\langle c_{li\sigma} c^{\dagger}_{l'j\sigma}\rangle c^{\dagger}_{li\sigma} c_{l'j\sigma}\nonumber\\
&=&\sum_{ll'k\sigma}\Sigma_{ll'}(k)c^{\dagger}_{lk\sigma}c_{l'k\sigma}\label {mfa1}
\end{eqnarray} 
where the self-energy $\Sigma_{ll'}(k)$ is defined by
\begin{eqnarray}
\Sigma_{ll'}(k) &=& \sum_{\vec d\ne 0}v_{li,l'j}\langle c_{li\sigma} c^{\dagger}_{l'j\sigma}\rangle \exp(i\vec k\cdot\vec d) \nonumber\\
&\equiv&\sum_{\vec d\ne 0}v_{li,l'j} [R_{ll'}(d)+iI_{ll'}(\vec d)]\exp(i\vec k\cdot\vec d) \label {mfa2}
\end{eqnarray} 
and $\vec d$ is the vector from the position $li$ to $l'j$. First, we consider the diagonal self-energy and denote $v_{li,lj}$ by $v(d)$ for brevity. Now, the function $R_{ll}(d)+iI_{ll}(\vec d)$ can be written as
\begin{eqnarray}
R_{ll}(d)+iI_{ll}(\vec d) &=& \langle c_{li\sigma} c^{\dagger}_{lj\sigma}\rangle\nonumber\\
&=& \frac{1}{2}(\langle c_{li\sigma} c^{\dagger}_{lj\sigma}\rangle -\langle c^{\dagger}_{lj\sigma}c_{li\sigma}\rangle)|_{i\ne j} \nonumber\\
&=&\frac{1}{N}\sum_k(\frac{1}{2}-\langle c^{\dagger}_{lk\sigma}c_{lk\sigma}\rangle )\exp(-i\vec k\cdot\vec d) \nonumber\\
\label {mfa3}
\end{eqnarray} 
where the $k$ summation runs over the first Brillouin zone, and $N$ is the total number of unit cells on single layer graphene. Note that the function $1/2-\langle c^{\dagger}_{lk\sigma}c_{lk\sigma}\rangle$ in the integrand in Eq. (\ref{mfa3})
is sizable only in areas close to the two Dirac points. Figure 3 shows the typical behaviors of the two functions 
\begin{figure} 
\centerline{\epsfig{file=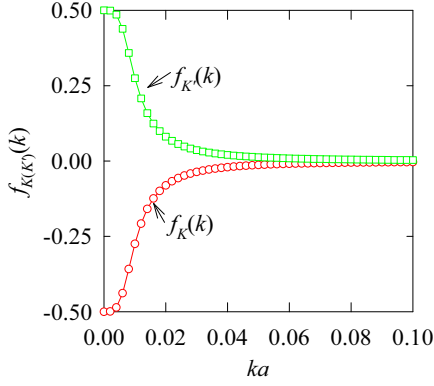,width=6. cm}}
\caption{(Color online) Behaviors of functions $f_K(k)$ (red circles) and $f_{K'}(k)$ (green squares) defined by Eq. (\ref{mfa4}). } 
\end{figure}
\begin{eqnarray}
f_{K(K')}(k)= \frac{1}{2}-\langle c^{\dagger}_{lK(K')+k\sigma}c_{lK(K')+k\sigma}\rangle|_{l=1} \label{mfa4}
\end{eqnarray} 
defined in the two valleys $K$ and $K'=-K$, respectively. The result in Fig. 3 is obtained by the self-consistent solution to the OCS without external fields. The functions are nonvanishing only within $k \leq 0.05/a$ with $a$ as the lattice constant. Then, the $k$ integration in Eq. (\ref{mfa3}) can be confined to two valleys. Since the range of the exchange interaction $v(d)$ is finite due to the electronic charge-fluctuation screening, the phase $\vec k\cdot\vec d$ in the factor $\exp(-i\vec k\cdot\vec d)$ can be safely approximated as $\vec K(\vec K')\cdot\vec d$. Therefore, we can write the formulas for $R_{ll}(d)$ and $I_{ll}(\vec d)$ as
\begin{eqnarray}
R_{ll}(d)&=& \frac{1}{N}{\sum_k}'(1-\langle c^{\dagger}_{lK+k\sigma}c_{lK+k\sigma}\rangle\nonumber\\
 & &~~~~~~~~ -\langle c^{\dagger}_{lK'+k\sigma}c_{lK'+k\sigma}\rangle)\cos(\vec K\cdot\vec d) \nonumber \\
&\equiv& r_{l} \cos(\vec K\cdot\vec d) \label{mfa5} \\
I_{ll}(\vec d)&=& \frac{1}{N}{\sum_k}'(\langle c^{\dagger}_{lK+k\sigma}c_{lK+k\sigma}\rangle \nonumber \\
& &~~~~~~~~-\langle c^{\dagger}_{lK'+k\sigma}c_{lK'+k\sigma}\rangle)\sin(\vec K\cdot\vec d)\nonumber \\
 &\equiv & -d_{l}\sin(\vec K\cdot\vec d) \label{mfa6} 
\end{eqnarray} 
where the $k$ summation is confined to a single valley, and the quantities $r_{l}$ and $d_{l}$ are defined by
\begin{eqnarray}
r_{l}&=& \frac{1}{N}{\sum_k}'(1-\langle c^{\dagger}_{lK+k\sigma}c_{lK+k\sigma}\rangle-\langle c^{\dagger}_{lK'+k\sigma}c_{lK'+k\sigma}\rangle) \nonumber\\
d_{l}&=& \frac{1}{N}{\sum_k}'(\langle c^{\dagger}_{lK'+k\sigma}c_{lK'+k\sigma}\rangle
-\langle c^{\dagger}_{lK+k\sigma}c_{lK+k\sigma}\rangle). \nonumber
\end{eqnarray} 
The quantity $r_l$ can be written as $r_{l} = -\delta_l/2$ with $\delta_l$ as the average electron doping concentration on sublattice $l$. For the doping concentration close to the CNP, we need to consider only the low energy quasiparticles with momenta close to the Dirac points. Then by expanding the self-energy with respect to the momentum $k$ in the two valleys and taking only the leading terms, we get 
\begin{eqnarray}
\Sigma_{ll}(\pm K)&=& r_{l}v_c \pm d_{l}v_s \nonumber\\
\end{eqnarray} 
with
\begin{eqnarray}
v_c&=& \sum_{\vec d\ne 0}v(d)\cos^2(\vec K\cdot\vec d) \nonumber\\
v_s&=& \sum_{\vec d\ne 0}v(d)\sin^2(\vec K\cdot\vec d). \label{mfa7}
\end{eqnarray} 

Physically, the imaginary part $I_{ll}(\vec d)$ is proportional to a bond current. All the bond currents in the lattice constitute to the current loops. The existence of the bond currents breaks the time-reversal symmetry. In Fig. 4, we draw out some of the bond currents on the same sublattice connected to a given site $i$. Clearly, the total current density at site $i$ is zero.

\begin{figure} 
\centerline{\epsfig{file=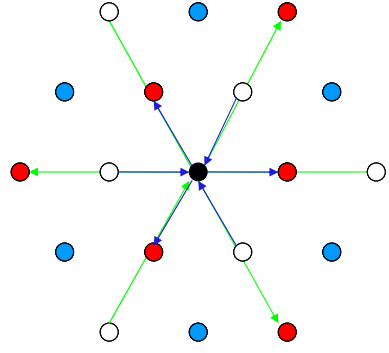,width=6 cm}}
\caption{(Color online) Some of the bond currents connected to the black site on a sublattice. The sign factor $\pm$ in $I_{ll}(\vec j-\vec i) \equiv \pm J_{l}(|\vec j-\vec i|)$ [as given by Eq. (\ref{mfa6})] is + (-) for the electron motion from the black site $i$ to the red (white) site $j$. There are no currents between the black site and the blue sites.} 
\end{figure}

Next, we consider the quantity $I_{ll'}(\vec d)$ with $l \ne l'$. For example, consider the case for $l$ = a$_1$ and $l'$ = b$_1$ on the top layer. Suppose the quantity is not vanishing. As shown in Fig. 5, the bond currents all with a fixed bond length $d = |\vec i-\vec j|$ result in three kinds of hexagon current loops with positive, negative, and zero fluxes [supposing the flux is positive (negative) for counterclockwise (clockwise) current loop], respectively. From the particle conservation law, the current along the boundary between the positive and the negative flux hexagons is two times of that along the boundary between the zero and the positive/negative flux ones. The hexagon current loops imply not only the breaking of time-reversal symmetry but also the breaking of translational invariance (homogeneity). The breaking of translational invariance to a low symmetry state requires the relevant interaction strong enough. Note that there is no a common periodicity for the two kinds of hexagon current loops with different side length $d$ in the lattice. The coexistence of the different hexagon current loops corresponds to completely an inhomogeneous system and cannot be realized for the electrons with finite-range interactions. The most favorable case is the smallest hexagon loops may exist when the interaction between the NN a$_1$ and b$_1$ atoms is strong enough. The argument applies to all $I_{ll'}(\vec d)$ with $l \ne l'$. For weak to medium interactions, we here assume all the currents between the sites of different sublattices are negligible small. On the other hand, the off-diagonal averages $\langle c_{li\sigma} c^{\dagger}_{l'j\sigma}\rangle$ with $l \ne l'$ can be pure real quantities without breaking homogeneity of the system. The real quantities describe the electron hopping and renormalize the noninteracting Hamiltonian. We here assume that such renormalization has already been included in $H_0$, we therefore do not take into account these hopping processes more again. (In the presence of external electric or magnetic field, even if the renormalization depends on the field, we will neglect the field effect.) 

\begin{figure} 
\centerline{\epsfig{file=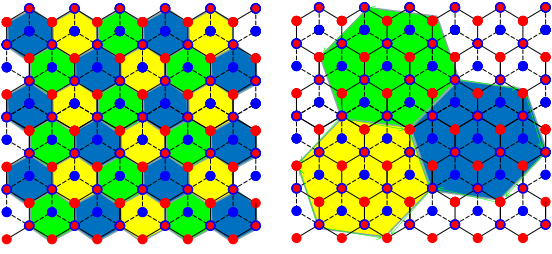,width=8 cm}}
\caption{(Color online) Two kinds of hexagon current loops on the top layer. Each bond of the hexagons connects the a$_1$ and b$_1$ atoms. The fluxes for the green, blue, and yellow hexagons are positive, negative, and zero, respectively.} 
\end{figure}

We suppose $d_{a_1}v_s=-d_{b_2}v_s \equiv-\Delta_1$ and $d_{b_1}v_s=-d_{a_2}v_s \equiv\Delta_2$ that means the breaking of the layer inversion symmetry. For the homogeneous system at the CNP, we have $r_l = 0$. As a result, the effective MFA Hamiltonian $H_{vk}$ is obtained by adding the diagonal matrix Diag($-s_v\Delta_1,s_v\Delta_2,-s_v\Delta_2,s_v\Delta_1$) to $H^0_{vk}$:
\begin{eqnarray}
H_{vk} = \begin{pmatrix}
-s_v\Delta_1& e_k&0&0\\
e^{\ast}_k&s_v\Delta_2&-t_1&0\\
0&-t_1&-s_v\Delta_2&e_k\\
0&0&e^{\ast}_k&s_v\Delta_1\\
\end{pmatrix}.\label{ocshk}
\end{eqnarray}
Note that the matrices $H_{vk}$ and $H_{-vk}$ are related by $H_{vk}=SH_{-v-k}S$, where $S$ is a $4\times 4$ matrix 
\begin{eqnarray}
S &=& \begin{pmatrix}
0& 0&0&1\\
0&0&1&0\\
0&1&0&0\\
1&0&0&0\\
\end{pmatrix}.\nonumber
\end{eqnarray}
If $\psi_k^{\mu}$ is an eigenfunction of $H_{vk}$ with eigenvalue $E_k^{\mu}$ (with $\mu$ = 1, 2, 3, 4), then $S\psi_{-k}^{\mu}$ is an eigenfunction of $H_{-vk}$ with the same eigenvalue. Therefore, the whole energy spectra can be obtained from the eigenstates only in a single valley.

\subsection {The OCS at $B$ = 0}

Under the MFA and with the wave functions $\psi_k^{\mu}$'s, the order parameters $\Delta_1$ and $\Delta_2$ are determined by
\begin{eqnarray}
\Delta_1&=& \frac{\sqrt{3}v_s}{2V}{\sum_{k\mu}}'f(E^{\mu}_{k})(|\psi_{k}^{1\mu}|^2-|\psi_{k}^{4\mu}|^2), \label{pd1}\\
\Delta_2 &=& \frac{\sqrt{3}v_s}{2V}{\sum_{k\mu}}'f(E^{\mu}_{k})(|\psi_{k}^{3\mu}|^2-|\psi_{k}^{2\mu}|^2), \label{pd2}
\end{eqnarray}
where $f$ is the Fermi distribution function, $\psi_k^{\nu\mu}$ is the $\nu$th component of the eigenfunction $\psi_k^{\mu}$, and $V=\sqrt{3}N/2$ is the total area of one layer.
From the 2BCM,\cite{Yan1} we know that the valence and conduction bands are connected to the electronic motions in the a$_1$ and b$_2$ sublattices. Therefore, the energy gap between the valence and conduction bands is determined by $2\Delta_1$. To reproduce the experimental data $|\Delta_1|$ = 1 meV at the CNP, $v_s$ needs to be $5.8\epsilon_0 = 14.06$ eV. Supposing the effective interaction \begin{eqnarray}
v(r) \approx e^2/\epsilon r[1+(\alpha r)^2]  
\end{eqnarray}
(decaying as $r^{-3}$, a typical behavior in the two-dimensional electron liquid \cite{Vignale}) with $\epsilon\approx 3$ as the screening constant of high frequency limit of BLG, we obtain the desired value $v_s=5.8\epsilon_0$ with $\alpha = 0.675$. Another coupling constant is obtained as $v_c \approx 4.7\epsilon_0$. Table I summaries all the parameters for the 4BCM.

\begin{table}[t]
\caption{\label{tab}Parameters for the 4BCM. }
\vspace{-3mm}
\begin{center}
\begin{ruledtabular}
\begin{tabular}{ccccccc}
$t$ (eV) & $t_1$ (eV)&$a$ (\AA)& $d$ (\AA)&$\alpha$ ($a^{-1}$)&$\epsilon$\\
\hline
2.8 & 0.39 &2.4 &3.34& 0.675 &3 \\
\end{tabular}
\end{ruledtabular}
\end{center}
\end{table}

\subsection {The OCS at finite $B$}

In the presence of the magnetic field $B$ applied perpendicularly to the sample plane, we take the Landau gauge for the vector potential, $\vec A = (0,Bx)$. With this gauge, the $y$ component momentum $k_y$ is a good quantum number. Replacing the variable $x$ and the operator $k_x = -i\nabla_x$ with the raising and lowering operators $a^{\dagger}$ and $a$, $k_y+Bx=\sqrt{B/2}(a^{\dagger}+a)$ and $k_x=i\sqrt{B/2}(a^{\dagger}-a)$, we can rewrite the effective Hamiltonian in real space. At the $K$ valley, the Hamiltonian is
\begin{eqnarray}
H_{Kx} = \begin{pmatrix}
-\Delta_1& i\sqrt{2B}a^{\dagger}&0&0\\
-i\sqrt{2B}a&\Delta_2&-t_1&0\\
0&-t_1&-\Delta_2&i\sqrt{2B}a^{\dagger}\\
0&0&-i\sqrt{2B}a&\Delta_1\\
\end{pmatrix}.\nonumber
\end{eqnarray}
Here $B$ is in the unit of $B_0 = \hbar c/ea^2 =1.105\times10^4$T. The $K$-valley eigenfunction $\psi^{\mu}_{Kn}$ is expressed as
\begin{equation}
\psi^{\mu}_{Kn} = (ix_{n}^{1\mu}\phi_n,
x_{n}^{2\mu}\phi_{n-1},
x_{n}^{3\mu}\phi_{n-1},
-ix_{n}^{4\mu}\phi_{n-2} 
)^t\nonumber
\end{equation}
for $n \geq$ 2, where $\phi_n$ is the $n$th level wave function of a harmonic oscillator of frequency $2B$ and mass 1/2 centered at $x_c=-k_y/B$, and the superscript $t$ means the transpose of the vector. The vector $X^{\mu}_{Kn} = (x_{n}^{1\mu},x_{n}^{2\mu},x_{n}^{3\mu},x_{n}^{4\mu})^{t}$ and the eigenenergy $E^{\mu}_{Kn}$ are determined by
\begin{eqnarray}
H_{Kn}X^{\mu}_{Kn} = E^{\mu}_{Kn}X^{\mu}_{Kn}   \label{egnk1}
\end{eqnarray}
with
\begin{eqnarray}
H_{Kn}=\begin{pmatrix}
-\Delta_1& \sqrt{2Bn}&0&0\\
\sqrt{2Bn}&\Delta_2&-t_1&0\\
0&-t_1&-\Delta_2&\sqrt{2B(n-1)}\\
0&0&\sqrt{2B(n-1)}&\Delta_1\\
\end{pmatrix}.\nonumber
\end{eqnarray}
The vector $X^{\mu}_{Kn}$ is normalized to unity. For each $n \geq 2$, the four energy levels appear at the valence, conduction, and other two bands about $\pm t_1$ far from the zero energy, respectively. For $n$ = 1, there are only three states with $x_{1}^{4\mu} = 0$ and the other three components and eigenvalues are determined by the upper left 3$\times$3 matrix of $H_{K1}$. For $n$ = 0, we have only one state $X^{1t}_{K0} = (1, 0, 0, 0)$ and $E^1_{K0} = -\Delta_1$. 

At the $K'$ valley, the Hamiltonian is
\begin{eqnarray}
H_{K'x} = \begin{pmatrix}
\Delta_1& i\sqrt{2B}a&0&0\\
-i\sqrt{2B}a^{\dagger}&-\Delta_2&-t_1&0\\
0&-t_1&\Delta_2&i\sqrt{2B}a\\
0&0&-i\sqrt{2B}a^{\dagger}&-\Delta_1\\
\end{pmatrix}.\nonumber
\end{eqnarray}
Since the Hamiltonian has the symmetry $H_{K'x}=SH_{Kx}S|_{i \to -i}$, the eigenfunction $\psi^{\mu}_{K'n}$ is therefore given as $S\psi^{\mu\ast}_{Kn}$ with the same eigen value $E^{\mu}_{Kn}$.

\begin{figure} 
\centerline{\epsfig{file=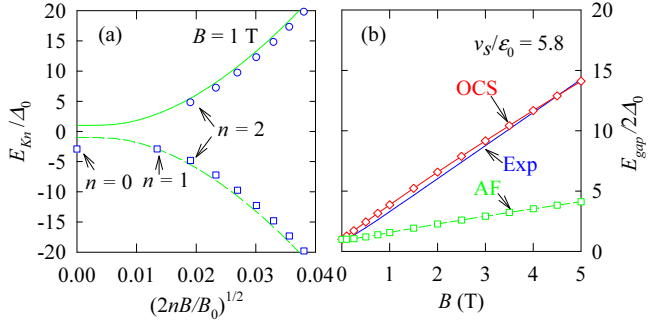,width=8.8 cm}}
\caption{(Color online) (a) Landau levels $E^{\mu}_{Kn}$ of the OCS in the valence (squares) and conduction (circles) bands at $B$ = 1 T. The lines represent the continuum conduction (solid) and valence (dashed) bands at $B = 0$ with momentum $k$ as the abscissa. (b) The gap $E_{\rm gap}$ (diamonds) as function of $B$ compared with the AF and the experimental results (Exp, Ref. \onlinecite{Velasco}).} 
\end{figure} 

In the presence of the magnetic field $B$, the formulas determining the order parameters $\Delta_{1,2}$ are different from Eqs. (\ref{pd1}) and (\ref{pd2}). The $k$ summations in Eqs. (\ref{pd1}) and (\ref{pd2}) are now replaced with the summations over $k_y$ and the Landau index $n$. Correspondingly, the wavefunction $\psi_{k}^{\nu\mu}/\sqrt{L_x}$ is replaced with $\psi_{Kn}^{\nu\mu}$ with $L_x$ as the length of the BLG in $x$ direction. By denoting the length in $y$ direction as $L_y$, we have $V = L_xL_y$. The $k_y$ summation is performed as
\begin{eqnarray}
\frac{1}{L_y}\sum_{k_y}|\psi_{Kn}^{\nu\mu}|^2&=& \frac{1}{2\pi}\int dk_y |\psi_{Kn}^{\nu\mu}|^2\nonumber\\
&=& \frac{B}{2\pi}\int dx_c|\psi_{Kn}^{\nu\mu}|^2\nonumber\\ 
&=& \frac{B}{2\pi}|x_{n}^{\nu\mu}|^2,  \label{kysum}
\end{eqnarray}
where $x_c$-integral has been carried out using the normalization condition for the wave functions of the harmonic oscillator. The equations for determining the order parameters are obtained as
\begin{eqnarray}
\Delta_1 &=& \frac{\sqrt{3}v_sB}{4\pi}\sum_{n\mu}f(E^{\mu}_{Kn})(|x_{n}^{1\mu}|^2-|x_{n}^{4\mu}|^2), \label{bd1}\\
\Delta_2 &=& \frac{\sqrt{3}v_sB}{4\pi}\sum_{n\mu}f(E^{\mu}_{Kn})(|x_{n}^{3\mu}|^2-|x_{n}^{2\mu}|^2). \label{bd2}
\end{eqnarray}

The solution to the Landau levels at $B$ = 1 T is shown in Fig. 6(a). Only the levels in the conduction and valence bands are depicted. For $n$ = 1, there is a level $E^v_{K1}$ slightly above $-\Delta_1(B)$ in the valence band. There is no state in the conduction band for $n$ = 0 and 1. Only when $n \geq 2$, the level $E^c_{Kn}$ in the conduction band appears. The energy gap is
\begin{eqnarray}
E_{\rm gap}=E^c_{K2}-E^v_{K1}.   \label{egap}
\end{eqnarray}
Clearly, the particle-hole symmetry is no longer valid at finite $B$, in agreement with the experiment.\cite{Velasco} Figure 6(b) shows $E_{\rm gap}$ of the OCS as function of $B$. The AF calculation of the same 4BCM (see Sec. IV) and experimental results for $E_{\rm gap}$ are also plotted for comparison. Here, the only fitting parameter is $v_s$ for reproducing $\Delta_1(0) = \Delta_0$ at $B$ = 0. The theoretical result for $E_{\rm gap}$ of the OCS as a function of $B$ is in surprisingly good agreement with the experiment.\cite{Velasco} 

\begin{figure} 
\centerline{\epsfig{file=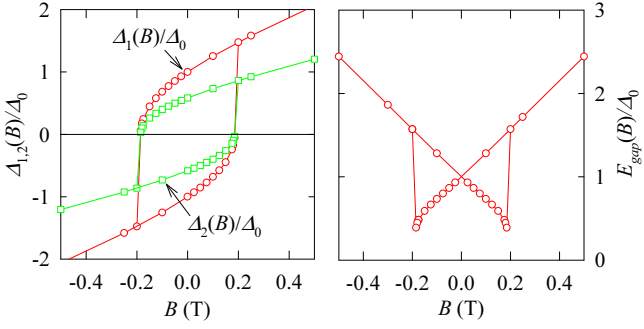,width=8.8 cm}}
\caption{(Color online) The hysteresis curves for $\Delta_1(B)$, $\Delta_2(B)$ (left) and $E_{\rm gap}(B)$ (right).} 
\end{figure} 

The above solution to the order parameters is only in the branch of $\Delta_{1,2} > 0$. At weak magnetic field $B > 0$, there is another branch of $\Delta_{1,2} < 0$. In this case, the two levels of $n$ = 0 and 1 appear in the conduction band but not in the valence band, and the energy gap is given by $E_{\rm gap}=E^c_{K1}-E^v_{K2}$. In Fig. 7, we show the hysteresis curves for the OCS order parameters $\Delta_1(B)$ and $\Delta_2(B)$ and the gap $E_{\rm gap}(B)$. For $|B|\leq 0.18$ T, there are two branches for $E_{\rm gap}$. In the lower gap branch, the gap decreases with increasing $|B|$. This behavior of $E_{\rm gap}$ is in qualitative agreement with the experimental observation by Weitz {\it et al.}\cite{Weitz} that indicates two peaks in the electric conductance appearing at $B_p \approx \pm 0.04$ T (where the real gap reaches the minimum), respectively.

\section{The AF state} \label{AF}

In the AF state, the magnetization at site $j$ is defined as
\begin{eqnarray}
m_j = \langle (n_{j\uparrow}- n_{j\downarrow})\rangle/2 
= -\langle \delta n_{j\downarrow}\rangle 
\end{eqnarray} 
where in the second equality we have used the facts that $\langle \delta (n_{j\uparrow}+ n_{j\downarrow})\rangle =0$ and the total number of up-spin electrons coincides with that of down-spin electrons. The magnetizations in an unit cell are given by $(m_{a_1}, m_{b_1}, m_{a_2}, m_{b_2})\equiv(m_1, -m_2, m_2, -m_1)$. The order parameters are defined as $-U(m_{a_1}, m_{b_1}, m_{a_2}, m_{b_2})\equiv(-\Delta_1, \Delta_2, -\Delta_2, \Delta_1)$.

Under the MFA, the interaction Hamiltonian reads
\begin{eqnarray}
H' &=& U\sum_{lj}(\delta n_{lj\uparrow}\langle \delta n_{lj\downarrow}\rangle+\langle \delta n_{lj\uparrow}\rangle\delta n_{lj\downarrow})\nonumber\\
&+&\sum_{li\ne l'j\sigma}v_{li,l'j}\langle c_{li\sigma} c^{\dagger}_{l'j\sigma}\rangle c^{\dagger}_{li\sigma} c_{l'j\sigma}. \label {mf1}
\end{eqnarray} 
By supposing $\langle c_{li\sigma} c^{\dagger}_{l'j\sigma}\rangle$ is real, the second term in right hand side of Eq. (\ref{mf1}) then describes the electron hopping and is a renormalization of the noninteracting Hamiltonian. As aforementioned, we suppose such a renormalization has already been included in the noninteracting Hamiltonian; we will not take into account this exchange effect again. 

With the MFA, we obtain an effective Hamiltonian as 
\begin{eqnarray}
H_{k\sigma} = \begin{pmatrix}
-\sigma\Delta_1& e_k&0&0\\
e^{\ast}_k&\sigma\Delta_2&-t_1&0\\
0&-t_1&-\sigma\Delta_2&e_k\\
0&0&e^{\ast}_k&\sigma\Delta_1\\
\end{pmatrix}\label{hk}
\end{eqnarray}
where $\sigma$ = +1 (-1) for spin-up (down) electrons. Note the matrices $H_{k\sigma}$ and $H_{k-\sigma}$ are related by
\begin{eqnarray}
H_{k-\sigma} &=& SH^{\ast}_{k\sigma}S. \nonumber
\end{eqnarray}
If $\psi_k^{\mu}$ is an eigenfunction of $H_{k\uparrow}$ with eigenvalue $E^{\mu}_{k}$ ($\mu$ = 1,2,3,4), then $S\psi_k^{\mu\ast}$ is an eigenfunction of $H_{k\downarrow}$ with the same eigenvalue. Therefore, we need to find out only the eigenstates of up-spin electrons.

\subsection{The AF state at $B$ = 0}

Using the property of the wave functions, we can obtain the equations for determining the order parameters. For $\Delta_1$, for example, we get
\begin{eqnarray}
\Delta_1 &=& \frac{U}{2N}\sum_{k\mu}f(E^{\mu}_{k})(|\psi_{k}^{1\mu}|^2-|\psi_{k}^{4\mu}|^2) \nonumber\\
&\approx& \frac{\sqrt{3}U}{2V}{\sum_{k\mu}}'f(E^{\mu}_{k})(|\psi_{k}^{1\mu}|^2-|\psi_{k}^{4\mu}|^2). \label{d1}
\end{eqnarray}
Here, the $k$ summation in the first line runs over the first Brillouin zone, while it runs over a single valley in the second line (because both valleys give the same contribution). Similarly, we obtain for $\Delta_2$,
\begin{eqnarray}
\Delta_2 \approx \frac{\sqrt{3}U}{2V}{\sum_{k\mu}}'f(E^{\mu}_{k})(|\psi_{k}^{3\mu}|^2-|\psi_{k}^{2\mu}|^2). \label{d2}
\end{eqnarray}

Equations (\ref{d1}) and (\ref{d2}) for determining the AF order parameters happen to be the same as Eqs. (\ref{pd1}) and (\ref{pd2}) for the OCS order parameters by setting $U = v_s$. Since the valence and conduction bands are connected to the electronic motions in the a$_1$ and b$_2$ sublattices, the energy gap between the valence and conduction bands is determined by $2\Delta_1$. To reproduce the experimental data $|\Delta_1|$ = 1 meV at the CNP, $U$ needs to be $5.8\epsilon_0 \approx 14.06$ eV. This value of U is larger than 9.3 eV of the recent {\it ab initio} calculation,\cite{Wehling} which means the AF state of $U = 9.3$ eV cannot reproduce the experimental data $\Delta_0$. 

\subsection{The AF state at finite $B$}

We now consider the behavior of the order parameters in the presence of the magnetic field $B$ applied perpendicularly to the BLG plane. Since the system under the magnetic field is not homogeneous, the Hamiltonian cannot be written in momentum space. For low energy electrons, however, their overall momenta are close to the Dirac points $K$ and $K'$. We here formulate the problem by a different way. From the beginning, we write the electron operator $c_{lj\sigma}$ as  
\begin{eqnarray}
c_{lj\sigma} = a^K_{lj\sigma}e^{i\vec K\cdot\vec j}+a^{K'}_{lj\sigma}e^{i\vec K'\cdot\vec j} \label{opa}
\end{eqnarray}
where $a^{K(K')}_{lj\sigma}$ is a fermion operator in valley $K (K')$ separated from the fast phase factor $\exp[i\vec K(\vec K')\cdot\vec j]$ and annihilates electrons of valley $K (K')$ and spin $\sigma$ at site $j$ of $l$ sublattice. The operator $a^{K(K')}_{lj\sigma}$ weakly depends on coordinate $j$. For later use, we here define the operator
\begin{eqnarray}
A^{\dagger}_{vj\sigma}=(a^{v\dagger}_{a_1j\sigma},a^{v\dagger}_{b_1j\sigma},a^{v\dagger}_{a_2j\sigma },a^{v\dagger}_{b_2j\sigma})   \label{opa2}
\end{eqnarray}
where $v = K$ or $K'$ is the valley index. In the presence of $B$, as did in Sce. III, we take the Landau gauge for the vector potential $\vec A = (0, Bx)$ and use the raising and lowering operators $a^{\dagger}$ and $a$. We get the effective Hamiltonian for AF state as
\begin{eqnarray}
H_{eff} = \sum_{vj\sigma}A^{\dagger}_{vj\sigma}H_{vj\sigma}A_{vj\sigma}  \nonumber
\end{eqnarray}
with
\begin{eqnarray}
H_{Kj\sigma} = \begin{pmatrix}
-\sigma\Delta_1& i\sqrt{2B}a^{\dagger}&0&0\\
-i\sqrt{2B}a&\sigma\Delta_2&-t_1&0\\
0&-t_1&-\sigma\Delta_2&i\sqrt{2B}a^{\dagger}\\
0&0&-i\sqrt{2B}a&\sigma\Delta_1\\
\end{pmatrix}\nonumber
\end{eqnarray}
for electrons at $K$ valley, and 
\begin{eqnarray}
H_{K'j\sigma} = \begin{pmatrix}
-\sigma\Delta_1& i\sqrt{2B}a&0&0\\
-i\sqrt{2B}a^{\dagger}&\sigma\Delta_2&-t_1&0\\
0&-t_1&-\sigma\Delta_2&i\sqrt{2B}a\\
0&0&-i\sqrt{2B}a^{\dagger}&\sigma\Delta_1\\
\end{pmatrix}\nonumber
\end{eqnarray}
for electrons at $K'$ valley. The Hamiltonian satisfies the transformation $H_{K'j-\sigma} = SH_{Kj\sigma}S|_{i\to -i}$.

As mentioned above, we need to find out the eigenstates of up-spin electrons,
\begin{eqnarray}
H_{vj\uparrow}\psi^{\mu}_{vn}(j) = E^{\mu}_{vn}\psi^{\mu}_{vn}(j) 
\end{eqnarray}
for $\mu$ = 1,2,3,4, and $n$ = 0, 1, $\cdots$. For each index $n$, the four energy levels (if they exist) appear at the valence, conduction, and other two bands about $\pm t_1$ far from the zero energy, respectively. At $K$ valley, the eigenfunction is given by 
\begin{eqnarray}
\psi^{\mu}_{Kn}(j) = \begin{pmatrix}
ix_{Kn}^{1\mu}\phi_n(j)\\
x_{Kn}^{2\mu}\phi_{n-1}(j)\\
x_{Kn}^{3\mu}\phi_{n-1}(j)\\
-ix_{Kn}^{4\mu}\phi_{n-2}(j)\\
\end{pmatrix}\label{wf}
\end{eqnarray}
for $n \geq$ 2. The vector $X^{\mu}_{Kn} = (x_{Kn}^{1\mu},x_{Kn}^{2\mu},x_{Kn}^{3\mu},x_{Kn}^{4\mu})^{t}$ and the eigenenergy are determined by
\begin{eqnarray}
H_{Kn}X^{\mu}_{Kn} = E^{\mu}_{Kn}X^{\mu}_{Kn}   \label{egnk}
\end{eqnarray}
with
\begin{eqnarray}
H_{Kn}=\begin{pmatrix}
-\Delta_1& \sqrt{2Bn}&0&0\\
\sqrt{2Bn}&\Delta_2&-t_1&0\\
0&-t_1&-\Delta_2&\sqrt{2B(n-1)}\\
0&0&\sqrt{2B(n-1)}&\Delta_1\\
\end{pmatrix}.\nonumber
\end{eqnarray}
The vector $X^{\mu}_{Kn}$ is normalized to unity. For $n$ = 1, there are only three states with $x_{K1}^{4\mu} = 0$ and the other three components and eigenvalues are determined by the upper left 3$\times$3 matrix of $H_{K1}$. For $n$ = 0, we have only one state $X^{1t}_{K0} = (1, 0, 0, 0)$ and $E^1_{K0} = -\Delta_1$. Note that this energy level is close to a level of $n=1$. On the other hand, at valley $K'$, the eigenfunction is given by
\begin{eqnarray}
\psi^{\mu}_{K'n} = ( 
ix_{K'n}^{1\mu}\phi_{n-2},
x_{K'n}^{2\mu}\phi_{n-1},
x_{K'n}^{3\mu}\phi_{n-1},
-ix_{K'n}^{4\mu}\phi_n)^t 
\nonumber
\end{eqnarray}
for $n \geq 2$. The eigen equation reads
\begin{eqnarray}
H_{K'n}X^{\mu}_{K'n} = E^{\mu}_{K'n}X^{\mu}_{K'n} \label{egnk'}
\end{eqnarray}
with
\begin{eqnarray}
H_{K'n}=\begin{pmatrix}
-\Delta_1& \sqrt{2B(n-1)}&0&0\\
\sqrt{2B(n-1)}&\Delta_2&-t_1&0\\
0&-t_1&-\Delta_2&\sqrt{2Bn}\\
0&0&\sqrt{2Bn}&\Delta_1\\
\end{pmatrix}.\nonumber
\end{eqnarray}
For $n$ = 1, we have three states with $x_{K'1}^{1\mu} = 0$ and the other three components and the eigenvalues are determined by the lower right 3$\times$3 matrix of $H_{K'1}$. For $n$ = 0, we have only $X^{1 t}_{K'0} = (0, 0, 0, 1)$ and $E^{1}_{K'0} = \Delta_1$ (close to a level of $n =1$). 

\begin{figure} 
\centerline{\epsfig{file=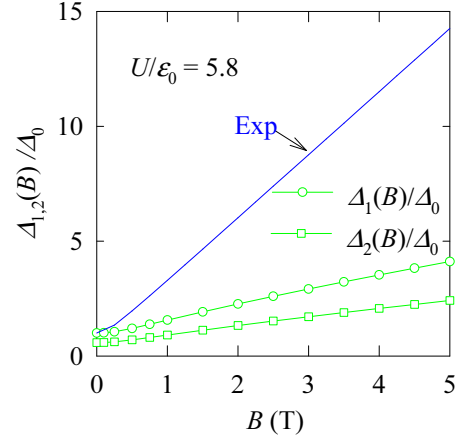,width=6.5 cm}}
\caption{(Color online) The AF order parameters $\Delta_1$ and $\Delta_2$ as functions of the magnetic field $B$. The blue solid line is the experimental result \cite{Velasco} for $E_{\rm gap}/2\Delta_0$.} 
\end{figure} 

The order parameter $\Delta_1$ is determined by
\begin{eqnarray}
\Delta_1 &=& \frac{U}{2}\sum_{v}(\langle a^{v\dagger}_{1j\uparrow}a^{v}_{1j\uparrow}\rangle- \langle a^{v\dagger}_{1j\downarrow}a^{v}_{1j\downarrow}\rangle)\nonumber\\
&=& \frac{\sqrt{3}U}{4L_y}\sum_{k_yvn\mu}f(E^{\mu}_{vn})(|\psi_{vn}^{1\mu}(j)|^2-|\psi_{vn}^{4\mu}(j)|^2) \nonumber\\
&=& \frac{\sqrt{3}UB}{8\pi}\sum_{vn\mu}f(E^{\mu}_{vn})(|x_{vn}^{1\mu}|^2-|x_{vn}^{4\mu}|^2) \label{ad1}
\end{eqnarray}
where the first line is the definition; the second line represents the averages in terms of the wave functions with $\psi_{vn}^{\nu\mu}$ as the $\nu$th component of $\psi_{vn}^{\mu}$, $S\psi^{\mu\ast}_{vn}$ has been used for spin down electrons, and a factor $\sqrt{3}/2$, the area of the unit cell of one layer graphene, comes from the fact that $|\psi_{vn}^{\nu\mu}(j)|^2/L_y$ is the probability density of electrons around site $j$ and the multiplication with $\sqrt{3}/2$ gives rise to the probability of electrons in the cell at site $j$; in the last line, the $k_y$ summation is carried out according to Eq. (\ref{kysum}). Analogously, the order parameter $\Delta_2$ is determined by
\begin{eqnarray}
\Delta_2 = \frac{\sqrt{3}UB}{8\pi}\sum_{vn\mu}f(E^{\mu}_{vn})(|x_{vn}^{3\mu}|^2-|x_{vn}^{2\mu}|^2). \label{ad2}
\end{eqnarray}

At the CNP and zero temperature, the order parameters $\Delta_1$ and $\Delta_2$ are self-consistently determined by Eqs. (\ref{egnk})-(\ref{ad2}). In Fig. 8, we show the results for $\Delta_1$ and $\Delta_2$ at zero temperature as functions of the magnetic field $B$ and compare them with the experimental data for $E_{\rm gap}/2\Delta_0$. Clearly, even though $\Delta_1$ and $\Delta_2$ grow with increasing $B$, their dependence of $B$ is not strong enough to match the experimental result.\cite{Velasco} Therefore we cannot expect the AF state as the candidate for the ground state of electrons in BLG.

\begin{figure} 
\centerline{\epsfig{file=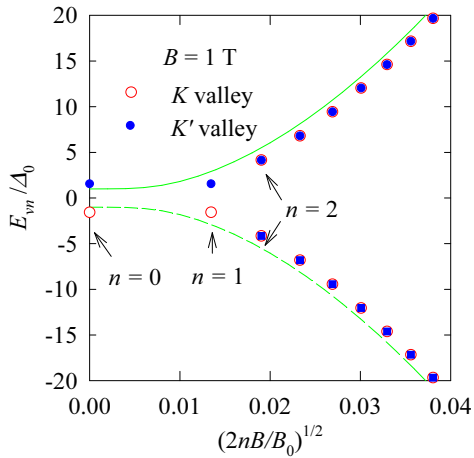,width=6.5 cm}}
\caption{(Color online) Landau levels $E^{\mu}_{vn}$ of the AF state in the valence and conduction bands at $B$ = 1 T. The circles and solid circles represent the levels in the $K$ and $K'$ valleys, respectively. The lines represent the continuum conduction (solid) and valence (dashed) bands at $B = 0$ with momentum $k$ as the abscissa. } 
\end{figure}

The Landau levels $E^{\mu}_{vn}$ of the AF state at $B = 1$ T are shown in Fig. 9. In different from the OCS, the distributions of the levels in the two valleys are now different. Especially, in the $K$ valley, there are no levels of $n$ = 0 and 1 in the conduction band (for positive $\Delta_{1,2}$), while they appear in the conduction band but disappear in the valence band in the $K'$ valley. The energy gap is therefore the indirect gap $E_{\rm gap} \approx  E^1_{K'0}-E^1_{K0} = 2\Delta_1$. 

As known, there is a momentum cutoff $k_c \approx a^{-1}$ for the 4BCM. The corresponding cutoff for the Landau levels is given by $n_c \approx B_0/2B$. At small $B$, $n_c$ is very large. For accelerating the numerical computation, we have used the super-high efficiency algorithm for sum of series.\cite{Yan3} According to the algorithm, one needs to compute only a number of selected Landau levels. 

\section{The OCS under external electric field} 

When an external electric field $E$ is applied perpendicularly to the BLG plane, there is an effective potential difference $2u=Eed/\epsilon$ between the two layers. The Hamiltonian $H_{vk}$ for the OCS now is obtained by adding the diagonal matrix Diag$(u+r_1v_c-s_v\Delta_1,u+r_2v_c+s_v\Delta_2,-u-r_2v_c-s_v\Delta_2,-u-r_1v_c+s_v\Delta_1)$ to $H^0_{vk}$. Here the terms $r_lv_c$ appear because of the electric polarization by $E$. Note that $r_l$ has the same sign of $u$, and thereby $H_{vk}(u)=SH_{-v-k}(-u)S$, which means that the order parameters are even functions of $u$. The model shows that if $E$ closes the energy gap, then $-E$ does it either. For the sake of illustration, we here consider the case of $B\geq 0$ and $u > 0$. The results for other cases can be deduced by the symmetry of the Hamiltonian. For $B\geq 0$ and $u > 0$, we still have two cases: $\Delta_1 > 0$ and $\Delta_1 < 0$. Here, we consider the case of $\Delta_1 > 0$. The discussion can be extended to the case of $\Delta_1 < 0$. At $B = 0$, the effective gap parameter is $u+r_1v_c-\Delta_1 \equiv E_{K0}$. The positive voltage $u$ pushes this level from the valence band toward to the conduction band. The critical potential $u_0$ closing the effective gap is obtained as $u_0 \approx 0.253 \Delta_0 \approx 0.253$ meV. The critical field of the experimental data \cite{Velasco} is $E \approx 1.25$ mV\AA$^{-1}$, which corresponds to $u_0 \approx 0.69$ meV (using $\epsilon \approx 3$). 

Since the system satisfies the particle-hole symmetry at $B = 0$, we can take the chemical potential as zero for the system at the CNP. Then, the level $E_{K0}$ is occupied if it is negative, otherwise it is empty. Therefore, with increasing $u$ from 0, the system undergoes a phase transition at $u=u_0$ from the state with the level $E_{K0}$ occupied to the state with the level empty. Thus, to search the critical $u_0$ where the gap closes at finite $B$, we study the phase transition.

\begin{figure} 
\centerline{\epsfig{file=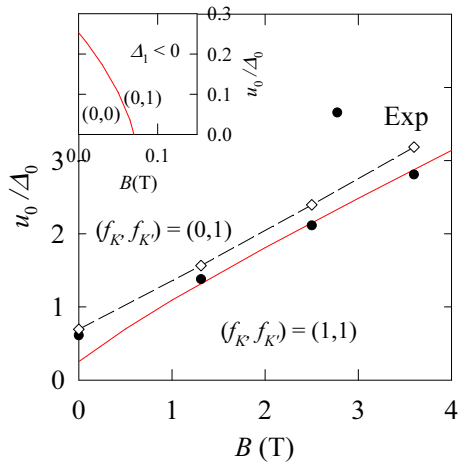,width=6.5 cm}}
\caption{(Color online) Phase boundary $u_0(B)$ between the two phases $(f_K,f_{K'}) = (0,1)$ and $(1,1)$ (the red-solid line). The solid points and the diamonds (connected by the dashed line) are converted from the critical $E$ of the experimental data (see Ref. \onlinecite{Velasco}) using $\epsilon \approx 3.4$ and 3, respectively. The inset shows the result for $\Delta_1 < 0$ in the range $0 < B < 0.15$ T.} 
\end{figure} 

At finite $B$, a state at $(B,u)$ can be obtained by continuously changing the parameters $B$ and $u$ from the state at $(0,u_i)$. If $u_i > u_0(0)$, then the level $E_{K0}$ is empty. Note that $E_{K0}$ is the only Landau level of $n$ = 0 at finite $B$ and there is another level of $n$ = 1 close to it similarly as the case of $u=0$. So the two levels of $n$ = 0 and 1 in the $K$ valley keep empty on the path from $(0,u_i)$ to $(B,u)$. On the other hand, if one starts from an initial state with $u_i < u_0(0)$, then the two levels of $n$ = 0 and 1 keep filled. (We denote the filling number as $f_K = 0$ and 1 for the two levels empty and filled, respectively.) We thus have two states at $(B,u)$. By comparing their energies, the ground state at $(B,u)$ is uniquely determined. At the critical potential $u_0(B)$, the two states have the same ground-state energy. The ground-state energy per unit cell, $E_0$, is given by
\begin{eqnarray}
E_0 = \frac{\sqrt{3}B}{4\pi}\sum_{vn\mu}f(E^{\mu}_{vn})[2E^{\mu}_{vn}-x_{vn}^{\mu\dagger}\Sigma(v)x_{vn}^{\mu}] \label{gse}
\end{eqnarray}
where $\Sigma(v)$ is the self-energy matrix given by
$\Sigma(v) = {\rm Diag}(r_1v_c-s_v\Delta_1,r_2v_c+s_v\Delta_2,-r_2v_c-s_v\Delta_2,-r_1v_c+s_v\Delta_1)$. The formula (\ref{gse}) can be derived according to many-particle theory.\cite{Fetter} 

Note that the energy levels of the OCS at finite $u$ are not degenerate for interchanging the indices of the two valleys. Especially, the Landau levels $E^c_{K0}$ and $E^c_{K'0}$ are given by $u+r_1v_c-\Delta_1$ and $-u-r_1v_c-\Delta_1$, respectively. For positive $u$ and $\Delta_1$, the level $E^c_{K'0}$ is always occupied.  

In Fig. 10, we exhibit the result for $u_0(B)$ as function of $B$ and compare it with the experimental data.\cite{Velasco} The experimental data are obtained by converting the critical electric field $E$ to $u_0$ according to $u_0 = Eed/2\epsilon$ with the dielectric constant $\epsilon \approx 3.4$ (solid points) and 3 (diamonds). As seen from Fig. 10, the behavior of $u_0(B)$ by the theoretical calculation is in fairly good agreement with the experiment \cite{Velasco} with $\epsilon \approx 3.4$ in the converting from $E$ to $u_0(B)$. 

As already seen, there is another solution of $\Delta_1 < 0$ in the range $0 < B < 0.18$ T. We show in the insert in Fig. 10 the phase boundary for this case. We see that the state of $\Delta_1 < 0$ in $B > 0.07$ T is unstable with respect to a small $E$. The range for the stable state of $\Delta_1 < 0$ is reduced to $|B| < 0.07$ T, with $|B_{max}| = 0.07$ T close toward to the experimental data \cite{Weitz} $|B_p| = 0.04$ T.

\section{Summary} 

With the MFA to the 4BCM, we have studied the OCS and the AF state of the electrons with finite-range repulsive interactions in BLG at the CNP. We have shown that the result of AF state is not in agreement with the experimental observation on the energy gap behavior that grows with increasing the magnetic field $B$. However, for the OCS with only one coupling constant $v_s$ fitting the experimental gap at $B$ = 0, the obtained energy gap at finite $B$ is in surprisingly good agreement with experimental data.\cite{Velasco} The results for the phase transition in the system in the presence of external electric and magnetic fields, and the particle-hole asymmetry spectra in the presence of $B$ are in qualitative agreements with the experimental observations.\cite{Velasco} There is also the intermediate experimental support \cite{Weitz} to the prediction for the hysteresis energy gap behavior with varying $B$. These facts show that the OCS is a possible ground state of electrons in BLG. The model explored here can be useful for understanding the physics of the electrons in BLG that is expected as a new generation of semiconductor. 

This work was supported by the National Basic Research 973 Program of China under Grants No. 2011CB932702 and No. 2012CB932302, NSFC under Grant No. 10834011, and the Robert A. Welch Foundation under Grant No. E-1146.

\end{document}